\documentclass[preprint,aps,showpacs,preprintnumbers,superscriptaddress,%
amsmath,amssymb]{revtex4}

\usepackage{graphicx}
\usepackage{dcolumn}
\usepackage{bm}

\begin{document}

\preprint{YITP-05-55}
\preprint{KUNS-1989}
\preprint{KEK-CP-171}
\preprint{CERN-PH-TH/2005-171}

\title{
  Topology conserving gauge action and the overlap-Dirac operator
}

\newcommand{\YITP}{
  Yukawa Institute for Theoretical Physics, 
  Kyoto University, Kyoto 606-8502, Japan
}
\newcommand{\CERN}{
  CERN, Theory Division, CH-1211 Geneva, Switzerland
}

\newcommand{\KUDP}{
  Department of Physics, 
  Kyoto University, Kyoto 606-8502, Japan
}

\newcommand{\GUAS}{
  School of High Energy Accelerator Science,
  The Graduate University for Advanced Studies (Sokendai),
  Tsukuba 305-0801, Japan
}

\newcommand{\KEK}{
  High Energy Accelerator Research Organization (KEK),
  Tsukuba 305-0801, Japan
}

\author{Hidenori~Fukaya}
\affiliation{\YITP}
\affiliation{\CERN}
\author{Shoji~Hashimoto}
\affiliation{\KEK}
\affiliation{\GUAS}

\author{Takuya~Hirohashi}
\affiliation{\KUDP}

\author{Kenji~Ogawa}
\affiliation{\GUAS}

\author{Tetsuya~Onogi}
\affiliation{\YITP}
\affiliation{\CERN}

\begin{abstract}
  We apply the topology conserving gauge action proposed by
  L\"uscher to the four-dimensional lattice QCD simulation
  in the quenched approximation. 
  With this gauge action the topological charge is
  stabilized along the hybrid Monte Carlo updates compared
  to the standard Wilson gauge action.
  The quark potential and renormalized coupling constant 
  are in good agreement with the results obtained
  with the Wilson gauge action.
  We also investigate the low-lying eigenvalue distribution
  of the hermitian Wilson-Dirac operator, which is relevant
  for the construction of the overlap-Dirac operator.
\end{abstract}

\maketitle

\section{Introduction}
Chiral symmetry and topology are tightly related with each
other in the gauge field theory through the quantum
correction. 
Namely, the axial anomaly appears at the one-loop level, and 
its integral over space-time leads to the topological charge
of the background gauge field.
In principle, one should be able to analyze the implication
of this relation for physical observables, such as the neutron
electric dipole moment, using the lattice gauge theory,
which provides a rigorous formulation of the non-Abelian
gauge theories even in the non-perturbative regime.
Such study is very difficult with the
Wilson-type Dirac operator, since the chiral symmetry is
explicitly violated on the lattice.


The overlap-Dirac operator
\cite{Neuberger:1997fp,Neuberger:1998wv} 
\begin{equation}
  \label{eq:ov}
  D = \frac{1}{\bar{a}} 
  \left[ 1 + \gamma_5 \mathrm{sgn}(aH_W) \right],
  \;\;\;
  \bar{a} = \frac{a}{1+s},
\end{equation}
realizes the exact chiral symmetry 
at finite lattice spacing $a$ \cite{Luscher:1998pq}
satisfying the Ginsparg-Wilson relation 
\cite{Ginsparg:1981bj} 
\begin{equation}
  \label{eq:GW_relation}
  \gamma_5 D+ D\gamma_5=aD \gamma_5 D.
\end{equation}
It is constructed from the Wilson-Dirac operator $aD_W$ with
the Wilson parameter $r=1$;
the hermitian Wilson-Dirac operator
$aH_W=\gamma_5(aD_W-1-s)$ 
enters as an argument of the sign function
$\mathrm{sgn}(x)$.
The parameter $s$ in (\ref{eq:ov}) is a fixed number in
the region $|s|<1$.

Since the definition (\ref{eq:ov}) contains a non-smooth
function, the locality of the Dirac operator could be lost
when there are near-zero eigenvalues of $|aH_W|$.
This is consistent with the index theorem, because the index
of the Dirac operator, which may be considered as a
definition of the topological charge, is a non-smooth
function of the background gauge field.
When the topological charge changes the value, the Dirac
operator must become ill-defined, and this is exactly the
point where $aH_W$ has a zero eigenvalue.

The locality of the overlap-Dirac operator (\ref{eq:ov})
is guaranteed for the gauge fields on which the minimum
eigenvalue of $|aH_W|$ is bounded from below by a positive
(non-zero) constant \cite{Hernandez:1998et}.
This condition is proved to be satisfied if the gauge field
configuration is smooth and each plaquette is close enough
to one; 
\begin{equation}
  \label{eq:admi}
  ||1-P_{\mu\nu}(x)|| < \epsilon \;\;\;\;
  \mbox{for all $x$, $(\mu,\nu)$}.
\end{equation}
Here, $P_{\mu\nu}(x)$ is the plaquette variable at $x$ on
the $\mu$-$\nu$ plane, and 
$||\cdots||$ denotes the norm of the operator.
In the four-dimensional case, the parameter 
$\epsilon\simeq 1/20.49$
is a sufficient (but not a necessary) condition
\cite{Neuberger:1999pz}.
This is called the ``admissibility'' bound.

One can construct a gauge action, which generates gauge
configurations respecting the condition (\ref{eq:admi}).
For instance, L\"uscher proposed the action
\cite{Luscher:1998du} 
\begin{equation}
  \label{eq:admiaction}
  S_G = \left\{
    \begin{array}{ll}\displaystyle
      \beta\sum_{P}\frac{1-\mbox{ReTr}P_{\mu\nu}(x)/3}
      {1-(1-\mbox{ReTr}P_{\mu\nu}(x)/3)/\epsilon},
      & \;\;\; \mbox{when} \;\;
      1-\mbox{ReTr}P_{\mu\nu}(x)/3 < \epsilon,
      \\
      \infty & \;\;\; \mbox{otherwise}
    \end{array}
    \right.,
\end{equation}
which has the same continuum limit as the standard Wilson
gauge action does.
In fact, the limit $\epsilon=\infty$ corresponds to the
standard Wilson gauge action.
Unfortunately, the bound $\epsilon\simeq 1/20.49$ is
too tight to produce gauge field ensembles corresponding to
the lattice spacing around 0.1~fm; 
for practical purposes, one must choose much larger values
of $\epsilon$.  
An interesting question is, then, whether the action can
keep the good properties for $\epsilon$ significantly
larger than $1/20.49$.
To be explicit, one expects that 
(i) the topology change during the molecular-dynamics-type
simulation is suppressed, 
and
(ii) the appearance of the near-zero eigenvalue of $|aH_W|$
is suppressed,
compared to the standard Wilson gauge action.
The point (i) is important in order to efficiently
generate gauge configurations with large topological
charge, which is necessary for the study of the
$\epsilon$-regime or the $\theta$-vacuum.
With the point (ii), the locality of the overlap-Dirac
operator is improved, and the numerical cost to apply the 
overlap-Dirac operator is reduced. 
In the numerical application to the massive Schwinger model 
with $\epsilon=1$, the stability of the topological charge
and the improvement of the chiral symmetry with the
domain-wall fermion were observed
\cite{Fukaya:2003ph,Fukaya:2004kp}.  
Also, in the four-dimensional quenched QCD, good stability
of the topological charge has been reported 
\cite{Shcheredin:2004xa,Bietenholz:2004mq,Shcheredin:2005ma,Nagai:2005fz}.

For gauge actions to be useful in practical simulations,
good scaling property toward the continuum limit is
required. 
The action (\ref{eq:admiaction}) differs from the standard
Wilson gauge action only at $O(a^4)$, and we expect that it
approaches to the continuum limit as quickly as the standard
action does.
The scaling would be better for the improved gauge actions,
such as the L\"uscher-Weisz \cite{Luscher:1985zq}, Iwasaki
\cite{Iwasaki:1985we} or DBW2 \cite{deForcrand:1999bi} gauge
actions, but 
an advantage of (\ref{eq:admiaction}) is that it contains a
parameter which directly controls the admissibility bound
and thus the appearance of the low-lying modes of $aH_W$.
For the improved actions including the rectangle loop, on
the other hand, the low-lying modes are suppressed for large
values of the rectangle coupling ({\it e.g.} with the DBW2
action) at the price of loosing the good scaling for
short distance quantities \cite{DeGrand:2002vu}. 

The goal of this paper is to give a systematic quenched QCD study of
the topology conserving gauge action. 
We find that the topology change is indeed suppressed when the
parameter $\epsilon$ is of order one. We also find that the scaling
violation in the static quark potential remain reasonably small 
and the tadpole improved perturbation theory for the renormalized 
gauge coupling shows a good convergence in the parameter range of our 
study. Therefore, the topology conserving gauge action has a desired 
properties for a practical application.  
Here we would like to comment on the possible application 
for the future work with the dynamical overlap fermion.
In the standard method, with the Wilson plaquette action,
one projects out the smallest eigenmodes of $H_W$
at every molecular dynamics step in the simulation trajectory 
and judge if the topology change occurs or not. 
When the topology change occurs, one recalculates 
the link updation with much higher accuracy on the topology crossing
point to choose either entering the new sector (refraction) or 
going back to the previous sector (reflection)~\cite{Fodor:2003bh}.
If one omits this step it would make the acceptance very low
due to the non-smoothness of the determinant as a functional 
of the gauge configuration.
On the other hand, if one uses the topology conserving action, 
crossing $H_W=0$ can be strongly suppressed,
so that one can avoid the CPU time consuming 
reflection/refraction method. 
In this sense, the use of the topology conserving gauge action can
also be useful in full QCD simulations with the overlap fermions.
Combination with the stout link version of 
the overlap fermion~\cite{Morningstar:2003gk} 
would be interesting as well.


This paper is organized as follows.
After describing the simulation methods in
Section~\ref{sec:simulation}, 
we show the fundamental scaling studies in
Section~\ref{sec:scaling}, that is the 
determination of the lattice spacing and a scaling test with
the static quark potential.
Renormalization of the coupling constant with the action
(\ref{eq:admiaction}) can be estimated using perturbation
theory as described in Section~\ref{sec:pert}.
Section~\ref{sec:Qstab} is the main part of this paper; we
report how much the topology change may occur with different
choices of parameters.
In Section~\ref{sec:overlap} the locality and the numerical
costs of the overlap fermion with gauge fields satisfying
the bound (\ref{eq:admi}) are discussed.
Conclusion and outlook are given in
Section~\ref{sec:conclusion}.

\section{Lattice simulations}
\label{sec:simulation}
Although several types of the gauge action that generate
the ``admissible'' gauge fields satisfying the bound 
(\ref{eq:admi}) are proposed
\cite{Shcheredin:2004xa,Bietenholz:2004mq},
we take the simplest choice (\ref{eq:admiaction}).
We study three values of $1/\epsilon$: 1, 2/3, and 0.
Note that $1/\epsilon=0$ corresponds to the conventional
Wilson gauge action.
The value $1/\epsilon = 2/3$ is the boundary,
below this value the gauge links can take any value in the 
gauge group $SU(3)$ and the positivity is guaranteed
\cite{Creutz:2004ir}.

The link variables are generated with the standard hybrid
Monte Carlo algorithm \cite{Duane:1987de}.
We take the molecular dynamics step size $\Delta \tau$ in
the range 0.01--0.02 and the number of steps in an unit
trajectory $N_{mds}$ = 20--40. 
During the molecular dynamics steps we monitor that the
condition $1-\mbox{ReTr}P_{\mu\nu}(x)/3 < \epsilon$ is
always satisfied with our choice of the step size.
We discarded at least 2000 trajectories for thermalization
before measuring observables.

\begin{figure}[tbp]
  \centering
  \includegraphics[width=8cm]{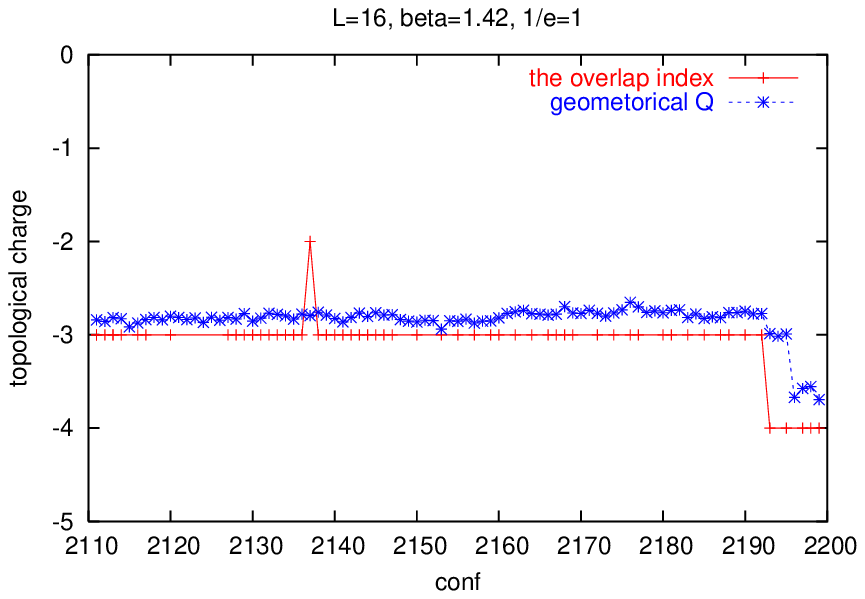}
  \includegraphics[width=8cm]{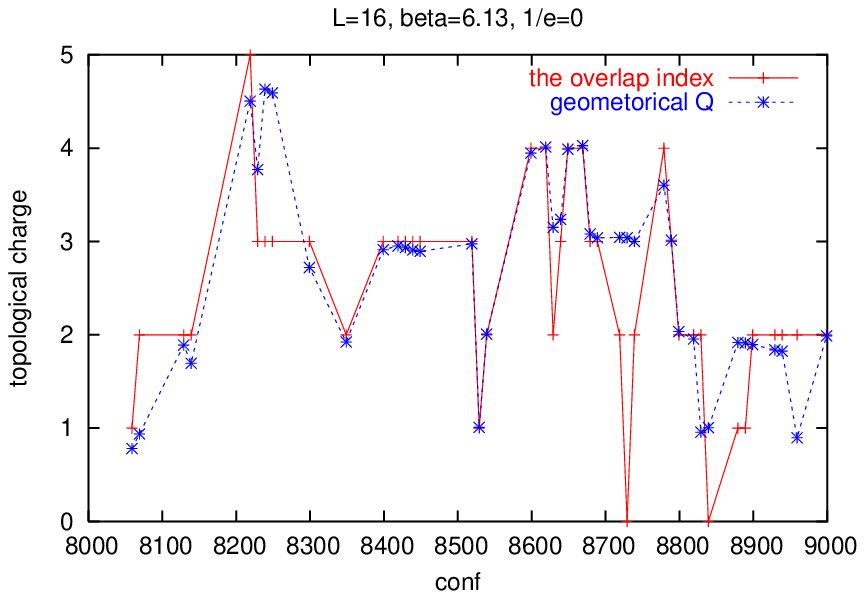}
  \caption{
    Comparison of the topological charge $Q$, calculated
    from the geometrical definition $Q_{geo}$
    (\ref{eq:topgeo}) as $Q=Z_{geo}Q_{geo}$, with the index
    of the overlap-Dirac operator. 
    $Q_{geo}$ is obtained by the cooling method (stars),
    and the index (pluses) is calculated for the overlap-Dirac
    operator with $s=0.6$. 
    The agreement is better for $1/\epsilon=1$ (left) than
    for $1/\epsilon=0$ (right).
  }
  \label{fig:Qhis}
\end{figure}

In order to measure the topological charge, we develop a new
type of cooling method.
It consists of the hybrid Monte Carlo simulation with an
exponentially increasing $\beta$ value
$\beta_{\mathrm{cool}}$ and decreasing step size
$\Delta\tau_{\mathrm{cool}}$ as a function of trajectory
$n_t$, {\it i.e.} 
\begin{eqnarray}
  \beta_{\mathrm{cool}}
  &=&\beta_{\mathrm{init}} \times(1.5)^{n_t},
  \nonumber\\
  \Delta\tau_{\mathrm{cool}} &=&
  \Delta\tau_{\mathrm{init}} \times(1.5)^{-n_t/2},
\end{eqnarray}
with a fixed $1/\epsilon_{\mathrm{cool}}$.
Note that 
$\sqrt{\beta_{\mathrm{cool}}}\Delta\tau_{\mathrm{cool}}$
is fixed so that the evolution at each step is kept small.
This method allows us to ``cool'' the configuration smoothly,
keeping the admissibility bound (\ref{eq:admi}) with
$1/\epsilon=1/\epsilon_{\mathrm{cool}}$. 
For the parameters, we take 
$(\beta_{\mathrm{init}},\Delta\tau_{\mathrm{init}},1/\epsilon_{\mathrm{cool}})$
= (2.0, 0.01, 1) for the configurations generated with
$1/\epsilon=1$, 
and (3.5, 0.01, 2/3) for the configurations with
$1/\epsilon=2/3$ or $1/\epsilon=0$. 
Even for the gauge configuration generated with the standard
gauge action ($1/\epsilon=0$), 
the condition $1/\epsilon_{\mathrm{cool}}=2/3$ can be used
because it allows all values of SU(3).
After 50--200 steps, the link variables are cooled down
close to a classical solution in each topological
sector. 
In fact, the geometrical definition of the topological charge
\cite{Hoek:1986nd}
\begin{equation}
  \label{eq:topgeo}
  Q_{geo}
  \equiv
  \frac{1}{32\pi^2}
  \sum_{x}\epsilon^{\mu\nu\rho\sigma} \mbox{ReTr}
  \left(P_{\mu\nu}(x)P_{\rho\sigma}(x)\right)
\end{equation}
of these ``cooled'' configurations gives numbers close to an
integer times a universal factor $Z_{geo}^{-1}$.
Namely, $Q=Z_{geo}Q_{geo}$ is close to an integer.
We determine $Z_{geo}$ through {\it would-be} $Q=1$ gauge
configurations, as $Z_{geo}^{-1}$ = 0.923(4).
As Figure~\ref{fig:Qhis} shows, the topological charge
$Q$ is consistent with the index of the overlap-Dirac
operator with $s$ = 0.6, which is calculated as described
in Section~\ref{sec:overlap}.
The consistency is better for $1/\epsilon=1$ than for the
standard Wilson gauge action $1/\epsilon=0$.

To generate topologically non-trivial gauge configurations,
we start the hybrid Monte Carlo simulation with the initial
condition 
\begin{eqnarray}
U_1(x)=\left(
\begin{array}{l}
e^{2\pi ix_4Q/L^2}\;\;\;\;\;\;\;\;\;\;\\
\;\;\;\;\;\;\;\;1\\
\;\;\;\;\;\;\;\;\;\;\;\;\;
e^{-2\pi ix_4Q/L^2}
\end{array}
\right),&
U_2(x)=\left(
\begin{array}{l}
1\;\;\;\;\;\;\;\;\;\;\\
\;\;\;\;\;\;\;\;
e^{2\pi i x_3 \delta_{x_2,L-1}/L}\\
\;\;\;\;\;\;\;\;\;\;\;\;\;\;\;\;\;
e^{-2\pi ix_3 \delta_{x_2,L-1}/L}
\end{array}
\right),\nonumber\\
U_3(x)=\left(
\begin{array}{l}
1\;\;\;\;\;\;\;\;\;\;\\
\;\;\;\;\;\;e^{-2\pi ix_2/L^2}\\
\;\;\;\;\;\;\;\;\;\;\;\;\;\;\;\;\;
e^{2\pi ix_2/L^2}
\end{array}
\right),&
U_4(x)=\left(
\begin{array}{l}
e^{-2\pi i x_1 Q\delta_{x_4,L-1}/L}\\
\;\;\;\;\;\;\;\;\;1\\
\;\;\;\;\;\;\;\;\;\;\;\;\;\;\;\;\;
e^{2\pi ix_1 Q\delta_{x_4,L-1}/L}
\end{array}
\right),
\end{eqnarray}
which is a discretized version of the classical solution on
a four-dimensional torus \cite{Gonzalez-Arroyo:1997uj}.
It can be used for any integer value of $Q$.
We confirmed that the topological charge assigned in this way
agrees with the index of the overlap operator with $s$ = 0.6.

We summarize the simulation parameters and the plaquette 
expectation values (for the run with the initial
configuration with $Q=0$) in Table~\ref{tab:param}.
The length of unit trajectory is 0.2--0.4, and the step size
is chosen such that the acceptance rate becomes larger than
$\sim$~70\%. 

\begin{table}[ptb]
\begin{center}
\begin{tabular}{ccccccc}
\hline\hline
Lattice size & $1/\epsilon$ & $\beta$ & $\Delta \tau$ & $N_{mds}$ &
acceptance & plaquette \\
\hline
$12^4$ & 1 & 1.0 & 0.01 & 40 & 89\% & 0.539127(9)\\
& & 1.2 & 0.01 & 40 & 90\% & 0.566429(6)\\
& & 1.3 & 0.01 & 40 & 90\% & 0.578405(6)\\
& 2/3 & 2.25 & 0.01 & 40 & 93\% & 0.55102(1)\\
&  & 2.4 & 0.01 & 40 & 93\% & 0.56861(1)\\
&  & 2.55 & 0.01 & 40 & 93\% & 0.58435(1)\\
& 0 & 5.8 & 0.02 & 20 & 69\% & 0.56763(5)\\
&  & 5.9 & 0.02 & 20 & 69\% & 0.58190(3)\\
&  & 6.0 & 0.02 & 20 & 68\% & 0.59364(2)\\
$16^4$ 
& 1 & 1.3 & 0.01 & 20 & 82\% & 0.57840(1)\\
& & 1.42 & 0.01 & 20 & 82\% & 0.59167(1)\\
& 2/3 & 2.55 & 0.01 & 20 & 88\% & 0.58428(2)\\
& & 2.7 & 0.01 & 20 & 87\% & 0.59862(1)\\
&0  & 6.0 & 0.01 & 20 & 89\% & 0.59382(5)\\
&  & 6.13 & 0.01 & 40 & 88\% & 0.60711(4)\\
$20^4$ 
& 1 & 1.3 & 0.01 & 20 & 72\% & 0.57847(9)\\
&  & 1.42 & 0.01 & 20 & 74\% & 0.59165(1)\\
& 2/3 & 2.55 & 0.01 & 20 & 82\% & 0.58438(2)\\
&  & 2.7 & 0.01 & 20 & 82\% & 0.59865(1)\\
& 0 & 6.0 & 0.015 & 20 & 53\% & 0.59382(4) \\
&  & 6.13 & 0.01& 20& 83\% & 0.60716(3)\\
\hline
\end{tabular}
\end{center}
\caption{
  Simulation parameters and the plaquette expectation values
  (for the run with the initial configuration with $Q=0$).
}
\label{tab:param}
\end{table}

\section{Static quark potential}
\label{sec:scaling}

In this section we describe the measurement of the static
quark potential to determine the lattice spacing for each
parameter choice.
We then compare the scaling violation and the rotational
symmetry violation with the case of the standard Wilson
gauge action. 
In the following, we assume that the topology of the gauge
field does not affect the Wilson loops, and choose the run
with $Q=0$ initial configuration for the measurement.

We measure the Wilson loops $W(\vec{r},t)$ 
using the smearing technique according to
\cite{Bali:1992ab}, 
where the spatial separation $\vec{r}/a$ is taken to be an
integer multiples of elementary vectors
$\vec{v}=(1,0,0)$, $(1,1,0)$,
$(2,1,0)$, $(1,1,1)$, $(2,1,1)$, $(2,2,1)$. 
With the assumption that the Wilson loop is an exponential 
function for large temporal side $t/a$,
$\langle W(\vec{r},t)\rangle =\exp(- V(\vec{r}) t)$,
we extract the static quark potential $a V(\vec{r})$.
The measurements are done every 20 trajectories and the
errors are estimated by the jackknife method.

As a reference scale, we measure the Sommer scales
$r_0$ and $r_c$ \cite{Guagnelli:1998ud,Necco:2001xg} defined
as $r_0^2 F(r_0)=1.65$ and $r_c^2 F(r_c)=0.65$, respectively.
Here, the force $F(r)$ on the lattice is given by a
derivative in the direction of $\vec{u}/a=(1,0,0)$;
\begin{equation}
  a^2 F(r_I)
  =
  \frac{aV(\vec{r})-aV(\vec{r}-\vec{u})}{|\vec{u}/a|},
\end{equation}
for $\vec{r}/|\vec{r}|=(1,0,0)$.
$r_I$ is introduced to cancel the discretization error in
the short distances, using the one-gluon exchange potential
on the lattice
\begin{eqnarray}
  \label{eq:r_I}
  \frac{1}{4\pi (r_I/a)^2}&=&
  -\frac{aG(\vec{r})-aG(\vec{r}-\vec{u})}{|\vec{u}/a|},
  \nonumber\\
  aG(\vec{r})&=&\int^{\pi}_{-\pi}\frac{d^3k}{(2\pi)^3}
  \frac{\prod^3_{j=1}\cos(r_jk_j/a) }{4\sum_{j=1}^3 \sin^2(k_j/2)}.
\end{eqnarray}

In Table~\ref{tab:sommerscale} we list the values of the
Sommer scales $r_0/a$, $r_c/a$ as well as their ratio
$r_c/r_0$.
The numerical results for $aV(\vec{r})$ and $r_I^2F(r_I)$ for
the case that $\vec{r}/a$ is an integer multiples of
$\vec{u}/a$ are given in 
Tables~\ref{tab:VF1} and \ref{tab:VF2}. 
The values of $r_I/a$ are also listed.

\begin{table}[tbp]
\begin{center}
\begin{tabular}{ccccccc}
\hline\hline
Lattice size & $1/\epsilon$ & $\beta$ & statistics & $r_0/a$ &
$r_c/a$ & $r_c/r_0$ \\
\hline
$12^4$ & 1 & 1.0 & 3800 & 3.257(30) & 1.7081(50)& 0.5244(52)\\
         & & 1.2 & 3800 &4.555(73)&2.319(10)&0.5091(81)\\
         & & 1.3 & 3800 &5.140(50)&2.710(14)&0.5272(53)\\
     & 2/3 & 2.25 & 3800 &3.498(24)&1.8304(60)&0.5233(41)\\
        &  & 2.4 & 3800 &4.386(53)&2.254(10)&0.5141(61)\\
        &  & 2.55 & 3800 & 5.433(72) & 2.809(18)&0.5170(67) \\
$16^4$ & 1 & 1.3 & 2300 & 5.240(96) & 2.686(13) & 0.5126(98) \\
         & & 1.42 & 2247 & 6.240(89) & 3.270(26)    &0.5241(83) \\ 
     & 2/3 & 2.55 & 1950 & 5.290(69) & 2.738(15) & 0.5174(72) \\
         & & 2.7 & 2150 & 6.559(76) & 3.382(22) & 0.5156(65) \\
\multicolumn{3}{l}{continuum limit \cite{Necco:2001xg}}
&&&& 0.5133(24)\\
\hline
\end{tabular}
\end{center}
\caption{
  Sommer scales $r_0/a$, $r_c/a$ and their ratio.
}\label{tab:sommerscale}
\end{table}

\begin{table}[tbp]
\renewcommand{\arraystretch}{0.9}
\begin{tabular}{ccc|cc|cc}
\hline\hline
& $1/\epsilon=1$& & \multicolumn{2}{c|}{12$^4$} & \multicolumn{2}{c}{16$^4$} \\
$\beta$ & $r/a$ & $r_I/a$ & $aV(\vec{r})$ & $r_I^2F(r_I)$ 
& $aV(\vec{r})$ & $r_I^2F(r_I)$ \\
\hline
1.0 & 1 &      & 0.50459(20) &            & & \\
    & 2 & 1.36 & 0.77828(61) & 0.5056(10) & & \\
    & 3 & 2.28 & 0.9629(15)  & 0.9520(69) & & \\
    & 4 & 3.31 & 1.1176(27)  &  1.691(26) & & \\
    & 5 & 4.36 & 1.2623(45)  &  2.751(80) & & \\
    & 6 & 5.39 & 1.4052(77)  &  4.33(22)  & & \\
\hline
1.2 & 1 &      & 0.44877(16) &            & & \\
    & 2 & 1.36 & 0.65982(39) & 0.38993(65)& & \\
    & 3 & 2.28 & 0.78291(80) & 0.6346(34) & & \\
    & 4 & 3.31 & 0.8775(13)  & 1.034(10)  & & \\
    & 5 & 4.36 & 0.9588(29)  & 1.545(45)  & & \\
    & 6 & 5.39 & 1.0322(47)  & 2.23(12)   & & \\
\hline
1.3 & 1 &      & 0.42730(10) &             & 0.42709(20) &            \\
    & 2 & 1.36 & 0.61711(34) & 0.35252(99) & 0.61710(66) & 0.35099(68)\\
    & 3 & 2.28 & 0.72140(69) & 0.53909(48) & 0.72130(92) & 0.5490(29) \\
    & 4 & 3.31 & 0.7977(12)  & 0.848(14)   & 0.7961(15)  & 0.8325(81) \\
    & 5 & 4.36 & 0.8608(21)  & 1.240(36)   & 0.8583(23)  & 1.180(32)  \\
    & 6 & 5.39 & 0.9230(25)  & 1.887(85)   & 0.9150(27)  & 1.809(79)  \\
    & 7 & 6.41 &             &             & 0.9636(51)  & 1.93(24)   \\
    & 8 & 7.43 &             &             & 1.0215(51)  & 3.09(37)   \\
\hline
1.42 & 1 &     &             &             & 0.40443(15) & \\
    & 2 & 1.36 &             &             & 0.57416(43) & 0.31444(58)\\
    & 3 & 2.28 &             &             & 0.66091(75) & 0.4567(22)\\
    & 4 & 3.31 &             &             & 0.7200(12) & 0.6583(61)\\
    & 5 & 4.36 &             &             & 0.7691(17) & 0.940(14)\\
    & 6 & 5.39 &             &             & 0.8076(24) & 1.189(48)\\
    & 7 & 6.41 &             &             & 0.8457(30) & 1.675(64)\\
    & 8 & 7.43 &             &             & 0.8832(37) & 1.91(14)\\
\hline
\end{tabular}
\caption{
  Potential and force values for the case that $\vec{r}/a$
  is an integer multiples of the unit vector
  $\vec{u}/a=(1,0,0)$. 
  Results for $1/\epsilon$ = 1.
}\label{tab:VF1}
\end{table}

\begin{table}[tbp]
\renewcommand{\arraystretch}{0.9}
\begin{tabular}{ccc|cc|cc}
\hline\hline
& $1/\epsilon = 2/3$
& & \multicolumn{2}{c|}{12$^4$} & \multicolumn{2}{c}{16$^4$} \\
$\beta$ & $r/a$ & $r_I/a$ & $aV(\vec{r})$ & $r_I^2F(r_I)$ 
& $aV(\vec{r})$ & $r_I^2F(r_I)$ \\
\hline
2.25 & 1 &      & 0.48470(15) & \\
     & 2 & 1.36 & 0.74012(57) & 0.47189(97) \\
     & 3 & 2.28 & 0.9077(13)  & 0.8640(56) \\
     & 4 & 3.31 & 1.0463(22)  & 1.515(21) \\
     & 5 & 4.36 & 1.1701(38)  & 2.353(64) \\
     & 6 & 5.39 & 1.2901(58)  & 3.64(15) \\
\hline
2.4 & 1 &      &   0.44908(12) & \\
    & 2 & 1.36 & 0.66434(41)   & 0.39770(70) \\
    & 3 & 2.28 & 0.79152(84)   & 0.6557(37) \\
    & 4 & 3.31 & 0.8889(15)    & 1.065(12) \\
    & 5 & 4.36 & 0.9749(23)    & 1.635(32) \\
    & 6 & 5.39 & 1.0541(30)    & 2.401(74) \\
\hline
2.55 & 1 &      & 0.42013(11) &             & 0.42042(16) &            \\
     & 2 & 1.36 & 0.60682(36) & 0.34493(58) & 0.60786(51) & 0.34590(72)\\
     & 3 & 2.28 & 0.70826(72) & 0.5230(28)  & 0.71227(95) & 0.5337(32) \\
     & 4 & 3.31 & 0.7806(13)  & 0.7913(90)  & 0.7878(16)  & 0.8211(93) \\
     & 5 & 4.36 & 0.8430(18)  & 1.187(18)   & 0.8538(22)  & 1.210(21)  \\
     & 6 & 5.39 & 0.8986(23)  & 1.686(37)   & 0.9157(29)  & 1.765(47)  \\
     & 7 & 6.41 &             &             & 0.9710(43)  & 2.229(84)  \\
     & 8 & 7.43 &             &             & 1.0266(52)  & 2.94(15)   \\
\hline
2.7 & 1.0 &      &            &             & 0.39590(15) & \\
    & 2.0 & 1.36 &            &             & 0.56100(44) & 0.30650(53)\\
    & 3.0 & 2.28 &            &             & 0.64733(62) & 0.4456(22)\\
    & 4.0 & 3.31 &            &             & 0.70527(90) & 0.6329(56)\\
    & 5.0 & 4.36 &            &             & 0.7528(14) & 0.907(14)\\
    & 6.0 & 5.39 &            &             & 0.7937(19) & 1.309(28)\\
    & 7.0 & 6.41 &            &             & 0.8321(24) & 1.531(44)\\
    & 8.0 & 7.43 &            &             & 0.8703(29) & 2.035(80)\\
\hline
\end{tabular}
\caption{
  Same as Table~\ref{tab:VF1}, but 
  for $1/\epsilon$ = 2/3.
}\label{tab:VF2}
\end{table}

\begin{figure}[tbp]
  \centering
  \includegraphics[width=12cm]{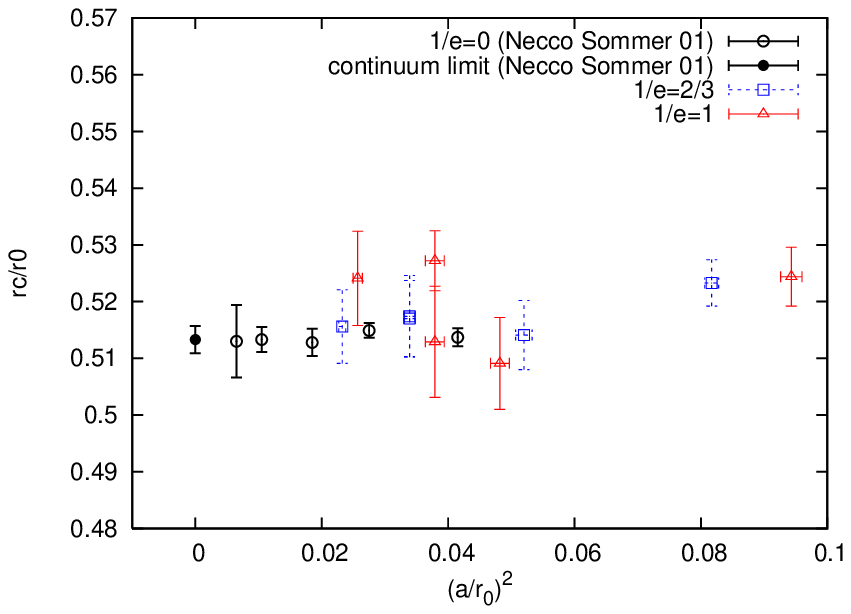}
  \caption{
    A ratio of the Sommer scales $r_c/r_0$.
    Squares and triangles are data for the topology
    conserving gauge action with $1/\epsilon$ = 2/3 and 1,
    respectively. 
    Open circles represent the standard Wilson
    gauge action (from \cite{Necco:2001xg}) and the filled
    circle is their continuum limit.
  }
  \label{fig:rcr0}
\end{figure}

The scaling can be tested for the ratio $r_c/r_0$.
Figure~\ref{fig:rcr0} presents the $a^2$ dependence of 
this ratio for different values of $1/\epsilon$.
Our results for $1/\epsilon$ = 2/3 and 1 are in perfect
agreement with the previous high statistics study for the
standard Wilson gauge action by Necco and Sommer
\cite{Necco:2001xg}.  
Moreover, we do not find any statistically significant
scaling violation except for the coarsest lattice points
around $(a/r_0)^2\simeq$ 0.1.

\begin{figure}[tbp]
  \centering
  \includegraphics[width=12cm]{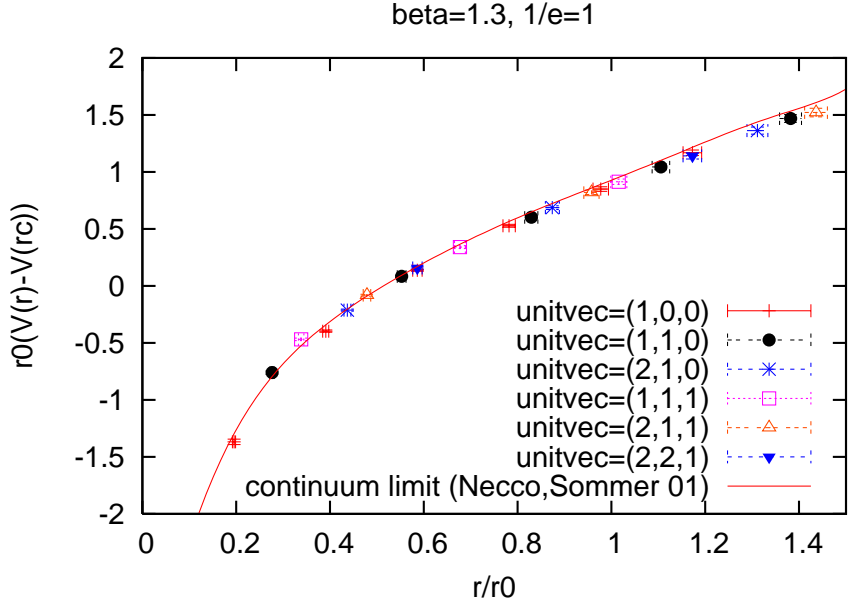}
  \caption{
    Static quark potential at $\beta=1.3$, $1/\epsilon$ = 1
    on a $12^4$ lattice. 
    The curve represents the continuum limit obtained by an
    interpolation of the results of \cite{Necco:2001xg}.
    Different symbols show $V(\vec{r})$'s with different
    orientations parallel to $\vec{v}$'s.
  }
  \label{fig:V(r)}
\end{figure}

Figure~\ref{fig:V(r)} shows a comparison of the potential
itself in a dimensionless combination, {\it i.e.}
$r_0\hat{V}(\vec{r})\equiv r_0(V(\vec{r})-V(r_c))$ versus
$|\vec{r}|/r_0$.
For $V(r_c)$ we interpolate the data 
in the direction $\vec{r}/r=(1,0,0)$.
The data at $\beta$ =1.3, $1/\epsilon=1$ are plotted
together with the curve representing the continuum limit
obtained in \cite{Necco:2001xg}.
The agreement is satisfactory (less than two sigma) for long
distances $r/r_0\gtrsim$ 0.5. 

\begin{figure}[thbp]
  \centering
  \includegraphics[width=12cm]{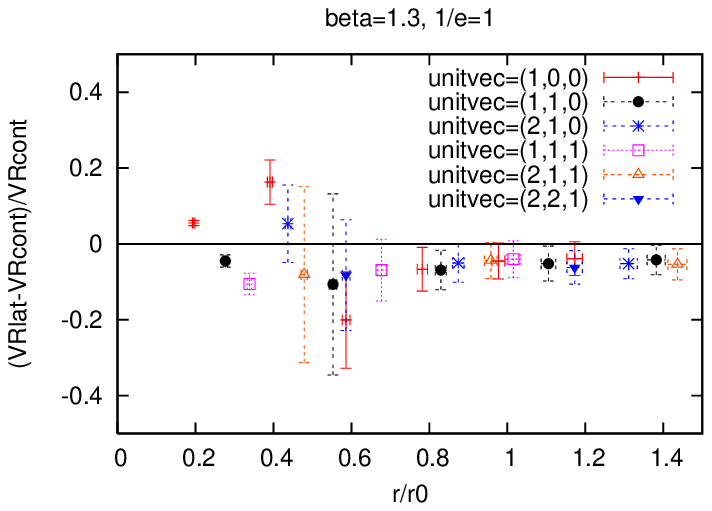}
  \caption{
    Violation of rotational symmetry 
    $(\hat{V}(\vec{r})-\hat{V}_{\mathrm{cont}}(|\vec{r}|))/%
    \hat{V}_{\mathrm{cont}}(|\vec{r}|)$,
    where $\hat{V}_{\mathrm{cont}}(|\vec{r}|)$ denotes the
    continuum limit.
    Results for $\beta$ = 1.3, $1/\epsilon$ = 1 are shown.
    The error of $\hat{V}_{\mathrm{cont}}(|\vec{r}|)$
    is not taken into account ($\lesssim 1\%$).\\\\
  }
  \label{fig:V(r)_rot}
  \centering
  \includegraphics[width=8cm]{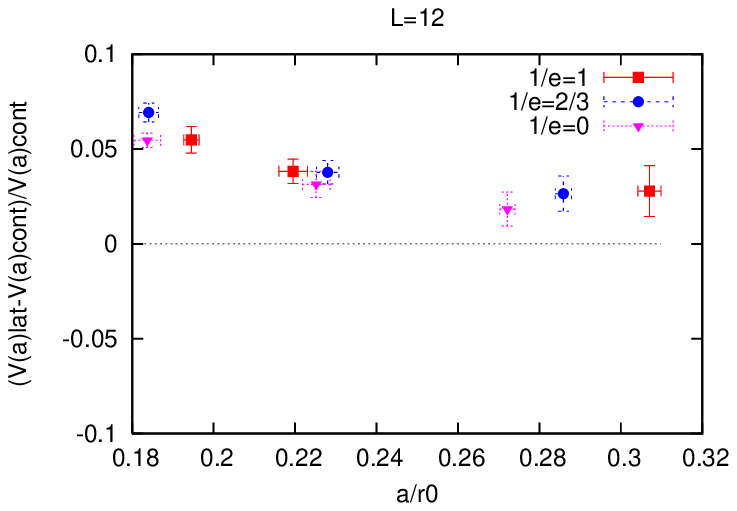}
  \includegraphics[width=8cm]{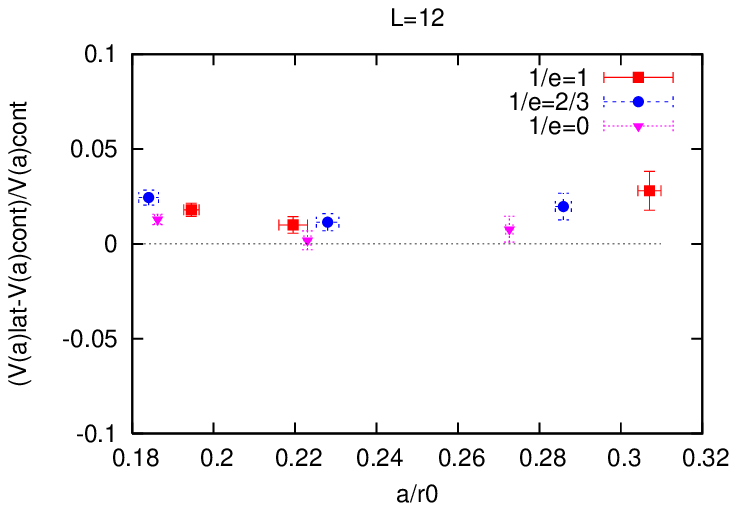}
  \caption{
    $(\hat{V}(\vec{r})-\hat{V}_{\mathrm{cont}}(|\vec{r}|))/
\hat{V}_{\mathrm{cont}}(|\vec{r}|)$ 
    for $\vec{r}=(1,0,0)$.
    Results with different $\beta$ and $1/\epsilon$ values
    are plotted as a function of $a/r_0$.
    Left and right panels show the plot without and with the
    tree level improvement for the argument $\vec{r}$,
    respectively.
    See the text for details.
  }
  \label{fig:rotation}
\end{figure}

For short distances, on the other hand, we can see
deviations of order 10\%, as shown in
Figure~\ref{fig:V(r)_rot}, where a ratio
$(\hat{V}(\vec{r})-\hat{V}_{\mathrm{cont}}(|\vec{r}|))
/\hat{V}_{\mathrm{cont}}(|\vec{r}|)$ is plotted.
$\hat{V}_{\mathrm{cont}}(|\vec{r}|)$ represents the curve in
the continuum limit drawn in Figure~\ref{fig:V(r)}.
The points corresponding to the separation $\vec{r}/a$ =
$(1,0,0)$ and $(2,0,0)$ deviates significantly from zero in the 
upward direction, while the points $(1,1,0)$ and $(1,1,1)$ are
lower than zero.
This implies the rotational symmetry violation.
Figure~\ref{fig:rotation} (left panel) shows the size of the
rotational symmetry violation at the point $(1,0,0)$ as a
function of the lattice spacing.
We find that the size of the violation is quite similar for
different values of $1/\epsilon$ including the standard
Wilson gauge action.
It does not show a tendency that the rotational symmetry
violation goes to zero in the continuum limit, but this
makes sense because the relevant scale of the observable is
also diverging as $1/a$.
After correcting the tree level violation by introducing
$d_I$ as $1/(4\pi d_I)=G(d)$, which is an analogue of $r_I$
in (\ref{eq:r_I}) but is defined for the potential, we
obtain the plot on the right panel of
Figure~\ref{fig:rotation}.
It is indeed improved.
The remaining correction is of order $\alpha_s(1/a)$, which
vanishes as $\sim 1/\ln(1/a)$ near the continuum limit.

These observations are consistent with the fact that the
topology conserving gauge action has the same $O(a^2)$
scaling violation as the standard Wilson gauge action.
The difference starts at $O(a^4)$, which is not visible at
the level of precision in our numerical study.

Finally, we confirm our assumption that the topology
does not affect the quark potential by measuring $r_0$ for
two initial value of $Q$ (0 and $-3$).
Measurements are done on a $16^4$ lattice at $\beta$ = 1.42,
$1/\epsilon$ = 1, for which the probability of the topology
change is extremely small as discussed in the next section.
Our results are $r_0/a$ = 6.24(9) for the $Q=0$ initial
condition and 6.11(13) for $Q=-3$.

\newpage

\section{Perturbative renormalization of the coupling}
\label{sec:pert}

In this section, we study whether the perturbative
corrections are under control with the topology conserving
gauge action.

Two-loop corrections to the gauge coupling for general
actions constructed by the plaquette is available in
\cite{Ellis:1983af}.
Using that formula, the renormalized gauge couping $g_{M}$
defined in the so-called Manton scheme is given by
\begin{eqnarray}
  \frac{1}{g_{M}^2}
  = \frac{1}{g^2} + A_1 + A_2 g^2,
\end{eqnarray}
where the coefficients $A_1$, $A_2$ are calculated as
\begin{eqnarray}
  A_1 & = &
  s_4\frac{2N_c^3-3}{N_c}
  +t_4(N_c^2+1),
  \nonumber\\
  A_2 & = &
  a_R [ s_4 (2N_c^2-3) + t_4 N_c (N_c^2+1) ]
  +s_6 \frac{15(N_c^4-3N_c^2+3)}{8N_c^2}
  \nonumber\\
  &&
  +u_6\frac{3(2N_c^2-3)(N_c^2+3)}{8N_c}
  +t_6\frac{3}{8}(N_c^2+1)(N_c^2+3)
  \nonumber\\
  && 
  -s_4^2 \frac{9N_c^4-30N_c^2+36}{2N_c^2}
  -2s_4 t_4 \frac{(2N_c^2-3)(N_c^2+2)}{N_c}
  -t_4^2 (N_c^2+1)(N_c^2+2).
\end{eqnarray}
Here, the parameters are $N_c=3$, 
$s_4=-1/4!$, 
$s_6=1/6!$, 
$t_4=1/(4N_c\epsilon)$,
$t_6=1/(8N_c^2\epsilon^2)$, 
$u_6=-1/(4!N_c\epsilon)$, and 
$a_R=-0.0011(2)$.
Table~\ref{tab:A1A2} gives the next-to-leading and
next-to-next-to-leading order coefficients $A_1$ and $A_2$
for various values of $\epsilon$. 

Since the perturbative expansion is poorly converging if one
uses the bare lattice coupling, we also consider the mean
field improvement using the measured value of the plaquette
expectation value \cite{Lepage:1992xa}.
To do so, we need a perturbative expectation value of the
plaquette expectation value, which is available to the
two-loop order for the general one-plaquette action
\cite{Heller:1995pu} as
\begin{eqnarray}
  \langle W(1,1)\rangle 
  &=&
  1 - g^2 \frac{(N_c^2-1)}{N_c}\bar{W}_2(1,1)
  - g^4 (N_c^2-1)X(1,1)\nonumber\\
  &&
  + g^4 \frac{(2N_c^2-3)(N_c^2-1)}{6N_c^2}\bar{W}_2(1,1)^2
  - g^4 \frac{(N_c^2-1)}{6N_c}C Z(1,1).
\end{eqnarray}
Here the notations $\bar{W}_2(1,1)$ and $X(1,1)$ are from the
original calculation \cite{Heller:1984hx} for the standard
Wilson gauge action, and 
$Z(1,1)=(1-1/V)\bar{W}_2(1,1)/4$ (on a symmetric lattice $V=L^4$)
is introduced for
generalization.
Their values are $\bar{W}_2(1,1)=1/8$, 
$X(1,1)=-1.01\times 10^{-4}$ and $Z(1,1)=1/32$ in the
infinite volume limit.
The constant $C$ is written as
\begin{equation}
  C = \left[\sum_R 6g^2 \frac{s_R(\beta)T(R)C_2(R)}{d_R}-N_c\right],
\end{equation}
where $C_2(R)$ is the quadratic Casimir operator in a
representation $R$ of the group $SU(N_c)$.
$d_R$ denotes the dimension of the representation $R$, and
$T(R)$ is defined such that 
$\mbox{Tr}_R(t^a t^b) = T(R)\delta^{ab}$
for the group generator $t^a$ in the representation $R$.
The coupling $s_R(\beta)$ is defined when we rewrite the gauge
action in terms of a general form of the one-plaquette action
\begin{equation}
  S_G = \sum_{x,\mu,\nu}\sum_R s_R(\beta)
  \left[
    1-\frac{1}{d_R} \mbox{Re}\mbox{Tr}_R P^R_{\mu\nu}(x)
  \right],
\end{equation}
where $P_{\mu\nu}^R$ denotes the plaquette in the $R$
representation.
The values of these parameters for the topology conserving
gauge action (\ref{eq:admiaction}) are
\begin{equation}
  s_3(\beta) = \left(1+\frac{11}{6\epsilon}\right)\beta,\;\;\;
  s_6(\beta) = -\frac{1}{3\epsilon}\beta,\;\;\;
  s_8(\beta) = -\frac{4}{9\epsilon}\beta,
\end{equation}
and 
$T(3)=1/2$, $T(6)=5/2$, $T(8)=3$,
$C_2(3)=4/3$, $C_2(6)=10/3$, $C_2(8)=3$.
Using these numbers, we obtain $C = 5-20/\epsilon$
and finally
\begin{equation}
  \langle W(1,1)\rangle
  = 1 - \frac{g^2}{3} +
  \left(\frac{5}{18\epsilon}-\frac{5}{144}\right)g^4.
\end{equation}

We define a boosted coupling $\bar{g}^2$ as
\begin{equation}
  \label{eq:boosted}
  \frac{1}{\bar{g}^2} = \frac{P}{g^2}
  \left(
    \frac{1}{1-(1-P)/\epsilon} +
    \frac{(1-P)/\epsilon}{(1-(1-P)/\epsilon)^2}
  \right),
\end{equation}
with the measured value of the plaquette expectation value
$P=\langle W(1,1)\rangle$ (see Table~\ref{tab:param}).
It is defined to be a factor in front of $F_{\mu\nu}^2$ when
we rewrite $P_{\mu\nu}=P\exp(ia^2F_{\mu\nu})$ and expand the
action (\ref{eq:admiaction}).
The perturbative expansion of (\ref{eq:boosted}) becomes
\begin{equation}
  \frac{1}{\bar{g}^2}
  = \frac{1}{g^2}+B_1 + B_2 g^2,
\end{equation}
where
\begin{equation}
  B_1 = -\frac{1}{3}
  \left(1-\frac{2}{\epsilon}\right),
  \;\;\;
  B_2 =
  \left(1-\frac{2}{\epsilon}\right)\left(
    \frac{5}{18\epsilon}-\frac{5}{144}
  \right)-\frac{2}{9\epsilon}+\frac{1}{3\epsilon^2}.
\end{equation}
We then obtain the perturbative expansion of the Manton
scheme coupling in terms of the boosted coupling
\begin{equation}
  \label{eq:g2conv_tad}
  \frac{1}{g_M^2} = 
  \frac{1}{\bar{g}^2}
  + (A_1-B_1) + (A_2-B_2) \bar{g}^2.
\end{equation}
Numerical values of $B_i$ and $A_i-B_i$ are listed in
Table~\ref{tab:A1A2}.
We can confirm the effect of the mean field improvement;
the two-loop coefficient $A_2$ is significantly reduced by
reorganizing the perturbative expansion as in
(\ref{eq:g2conv_tad}).

\begin{table}[tbp]
  \centering
  \begin{tabular}{ccccccc}
    \hline\hline
    $1/\epsilon$ & $A_1$ & $A_2$ & $B_1$ & $B_2$ &
    $A_1-B_1$ & $A_2-B_2$\\
    \hline
    0   & $-$0.20833 & $-$0.03056 
        & $-$0.33333 & $-$0.03472
        &    0.12500 &    0.00416\\
    2/3 &    0.34722 & $-$0.04783
        &    0.11111 & $-$0.05015
        &    0.23611 &    0.00233\\
    1   &    0.62500 & $-$0.10276
        &    0.33333 & $-$0.13194
        &    0.29167 &    0.02919\\
    \hline
  \end{tabular}
  \caption{
    Next-to-leading and next-to-next-to-leading order
    coefficients for the coupling renormalization for
    various $\epsilon$. 
    Mean field improved coefficients $A_1-B_1$, $A_2-B_2$ are
    also shown. 
    See the text for details.
  }
  \label{tab:A1A2}
\end{table}
\begin{figure}[bthp]
  \centering
  \includegraphics[width=9.5cm]{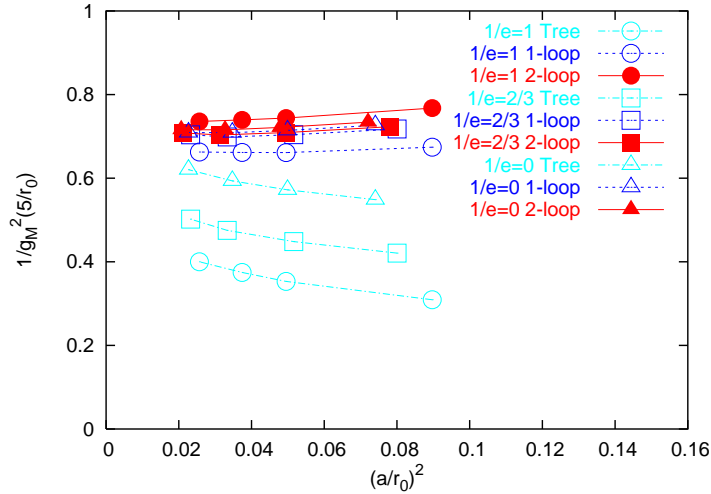}
  \caption{
    $1/g_M^2$ in the Manton scheme for various values of
    $1/\epsilon$.
    The mean field improved expression (\ref{eq:g2conv_tad}) is
    used with the measured plaquette expectation value.
    Different symbols distinguish the value of $1/\epsilon$
    (1 for circles, 2/3 for squares, 0 for triangles).
    Open symbols with dot-dashed line represent tree-level
    results and open symbols with dashed lines are one-loop.
    The best results including two-loop corrections are
    shown by filled symbols with solid lines.
  }
  \label{fig:manton}
\end{figure}

Using these results, the inverse squared renormalized
coupling in the Manton scheme is obtained for each lattice
parameters. 
In Figure~\ref{fig:manton} we plot the coupling evaluated at
a reference scale $5/r_0$ as a function of lattice scaling
squared. 
We use the two-loop renormalization equation for the
evolution to the reference scale.
Although the couplings are very different at the tree level,
the one-loop results are already in good agreement among the
different values of $1/\epsilon$.
Including the two-loop corrections, we find that the
perturbative expansion converges very well and the agreement
among different $1/\epsilon$ becomes even better.
Good scaling toward the continuum limit can also be observed
in this plot for the two-loop results.

\newpage
\section{Stability of the topological charge}
\label{sec:Qstab}

In this section we discuss the stability of the topological
charge with the topology conserving gauge action.

How the topological charge is preserved can be easily
explained in the U(1) gauge theory in two-dimension, for
which we can define an exact geometrical definition of the
topological charge \cite{Luscher:1998kn,Luscher:1998du}
\begin{eqnarray}
  Q_{\mathrm{geo}} &=& \frac{1}{2\pi}
  \sum_{x}\frac{1}{2}\epsilon_{\mu\nu}
  F^{\mathrm{lat}}_{\mu\nu}(x),\nonumber\\
  F^{\mathrm{lat}}_{\mu\nu}(x)&=&
  -i\ln (P_{\mu\nu}(x)),\;\;\; -\pi <
  F^{\mathrm{lat}}_{\mu\nu}(x)\leq \pi.
\end{eqnarray}
$P_{\mu\nu}(x)$ denotes the plaquette in the U(1) gauge theory.
In two dimensions, $Q_{\mathrm{geo}}$ gives an integer on
the lattices with the periodic boundary condition.
The topological charge may change its value when the field
strength pass through the point 
$F^{\mathrm{lat}}_{\mu\nu}(x) = \pm\pi$.
Since the jump from 
$F^{\mathrm{lat}}_{\mu\nu}(x)=-\pi$ to 
$F^{\mathrm{lat}}_{\mu\nu}(x)=+\pi$ is allowed with the
usual compact and non-compact gauge actions, the topology
change may occur without a big penalty.
It is the U(1) version of the L\"uscher's bound
\begin{equation}
  1-\mbox{Re} P_{\mu\nu}(x) < \epsilon
\end{equation}
with $\epsilon < 2$, that can prevent these topology
changes because the point 
$F^{\mathrm{lat}}_{\mu\nu}(x) = \pm\pi$
is not allowed under this condition.
Furthermore, it can be shown that $Q_{\mathrm{geo}}$ is
equivalent to the index of the overlap fermion with $s=0$ if
$\epsilon < 1/5$ is satisfied. 

For the non-abelian gauge theories in 
higher dimensions, we do not have the exact geometrical
definition of the topological charge
(note that (\ref{eq:topgeo}) gives non-integers).
It is, however, quite natural to assume 
that a similar mechanism concerning the compactness 
of the link variables allows us to preserve the index of 
the overlap-Dirac operator for very small $\epsilon$.
Also for larger $\epsilon$, we may expect that the topology 
stabilizes well in practical sampling of gauge
configurations. 

Table.~\ref{tab:Qstab} summarized our data for the stability
of the topological charge 
\begin{eqnarray}
  \mbox{Stab}_Q \equiv \frac{N_{\mathrm{trj}}}
  {\tau_{\mathrm{plaq}}\times \#Q},
\end{eqnarray}
where $\tau_{\mbox{\tiny plaq}}$ is the autocorrelation time
of the plaquette, measured using the method described in
Appendix E of \cite{Luscher:2004rx}.
$N_{\mbox{\tiny trj}}$ denotes the total length of the HMC
trajectories and 
$\#Q$ is the number of topology changes during the
trajectories.
The topological charge $Q$ is measured every 10--20 trajectories
with the geometrical definition (\ref{eq:topgeo}) after our
cooling method. 
With this definition, $\mbox{Stab}_Q$ represents a mean
number of independent gauge configurations sampled staying a
certain topological charge. 
But it only gives an upper limit, because the topology
change is detected only every 10--20 trajectories and we may
miss the change if $Q$ changes its value and returns to the
original value between two consecutive measurements.
Therefore, our measurement of $\mbox{Stab}_Q$ may give a
good approximation when the topology change is a rare
event.

\begin{table}[tbp]
\begin{center}
\begin{tabular}{cccccccc}
\hline\hline
Lattice size & $1/\epsilon$ & $\beta$ & $r_0/a$ &
$N_{\mbox{\tiny trj}}$ & $\tau_{\mbox{\tiny plaq}}$ & $\#Q$ & Stab$_Q$ \\
\hline
$12^4$ & 1   & 1.0  & 3.257(30) & 18000 & 2.91(33) & 696&9\\
           & & 1.2  & 4.555(73) & 18000 & 1.59(15) & 265 &43\\
           & & 1.3  & 5.140(50) & 18000 & 1.091(70) & 69& 239\\
       & 2/3 & 2.25 & 3.498(24) & 18000 & 5.35(79) & 673 & 5\\
          &  & 2.4  & 4.386(53) & 18000 & 2.62(23) & 400 & 17\\
          &  & 2.55 & 5.433(72) & 18000 & 2.86(33) & 123 & 51\\
       & 0   & 5.8  & [3.668(12)] & 18205 & 30.2(6.6) & 728 & 1\\
          &  & 5.9  & [4.483(17)] & 27116 & 13.2(1.5) & 761 & 3\\
          &  & 6.0  & [5.368(22)] & 27188 & 15.7(3.0) & 304 & 6\\
$16^4$ & 1   & 1.3  & 5.240(96) & 11600 & 3.2(6) & 78 & 46\\
           & & 1.42 & 6.240(89) & 5000 & 2.6(4) & 2 & 961\\ 
       & 2/3 & 2.55 & 5.290(69) & 12000 & 6.4(5) & 107 & 18\\
           & & 2.7  & 6.559(76) & 14000 & 3.1(3) & 6 & 752\\
       & 0   & 6.0  & [5.368(22)] & 3500 & 11.7(3.9) & 14 & 21\\ 
          &  & 6.13 & [6.642(--)] & 5500 & 12.4(3.3) & 22 & 20\\
$20^4$ & 1   & 1.3  & --- & 1240 & 2.6(5) & 14 & 34\\
          &  & 1.42 & --- & 7000 & 3.8(8) & 29 & 64 \\ 
       & 2/3 & 2.55 & --- & 1240 & 3.4(7) & 15 & 24\\
          &  & 2.7  & --- & 7800 & 3.5(6) & 20 & 111\\
       & 0   & 6.0  & --- & 1600 &14.4(7.8) & 37 & 3 \\
          &  & 6.13 & --- & 1298 & 9.3(2.8) &4  &35 \\
\hline
\end{tabular}
\end{center}
\caption{
  Stability of the topological charge $\mbox{Stab}_Q$.
  The length of the HMC trajectory $N_{\mbox{\tiny trj}}$,
  the autocorrelation time measured for plaquette
  $\tau_{\mbox{\tiny plaq}}$,
  and the number of topology change $\#Q$ are also
  summarized. 
  }
  \label{tab:Qstab}
\end{table}

\begin{figure}[tbp]
  \centering
  \includegraphics[width=10cm]{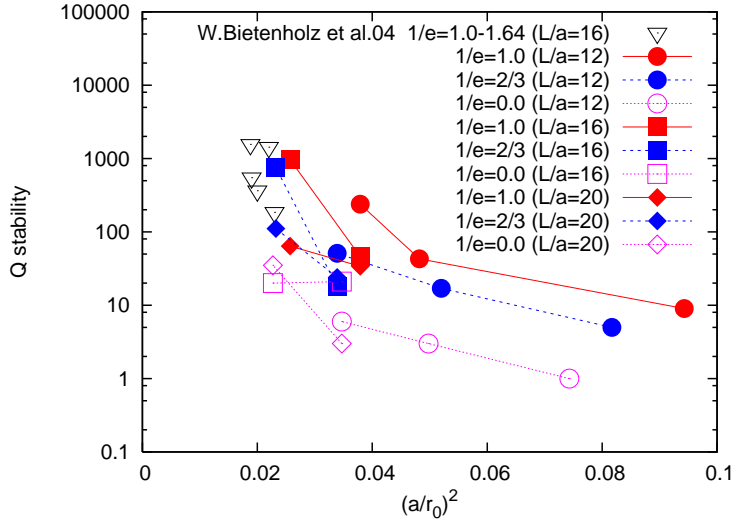}
  \caption{
    Stability of the topological charge.
    Different symbols represent the lattice size:
    $L/a$ = 12 (circles), 16 (squares), 20 (upward
    triangles). 
    The value of $1/\epsilon$ is distinguished by the line
    type:
    $1/\epsilon$ = 0 (dotted), 2/3 (dashed), 1 (solid).
    Downward triangles are the data of
    \cite{Bietenholz:2004mq} measured on 16$^4$ lattices. 
  }
  \label{fig:Qstab}
\end{figure}

Results are plotted in Figure~\ref{fig:Qstab} as a function
of the lattice spacing squared.
We find a clear trend that the stability increases for
larger $1/\epsilon$ if the lattice spacing is the same.
When the lattice size is increased from $L/a$ = 12 to 16,
the stability drops significantly for each value of
$1/\epsilon$.
This is expected, because the topology change occurs through
local dislocations of gauge field and its probability scales
as the volume.
For even larger volume ($L/a$ = 20), our data are not
precise enough, since the total length of trajectory is
shorter.
We also observe that the stability increases very rapidly
toward the continuum limit.

For the study of the $\epsilon$-regime in a fixed topological 
sector, the lattices
$(1/\epsilon, \beta, L)\sim (1,1.42,16)$ and 
$(2/3, 2.7, 16)$ would be appropriate.
Their physical size is $L \sim$ 1.25~fm and the topological
charge is stable for 
$(100-1000)\tau_{\mathrm{plaq}}$ trajectories.

\section{Construction of the overlap-Dirac operator}
\label{sec:overlap}

\subsection{Low-lying mode distribution of $H_W$}
We measure the low-lying eigenvalues of $aH_W$ on the gauge
configurations generated with the topology conserving gauge
action.
We use the numerical package ARPACK \cite{ARPACK}, which
implements the implicitly restarted Arnoldi method.
For the hermitian Wilson-Dirac operator $aH_W$ we take the
form $aH_W=\gamma_5(aD_W-1-s)$ with $s=0.6$.

\begin{figure}[btp]
  \centering
  \includegraphics[width=10cm]{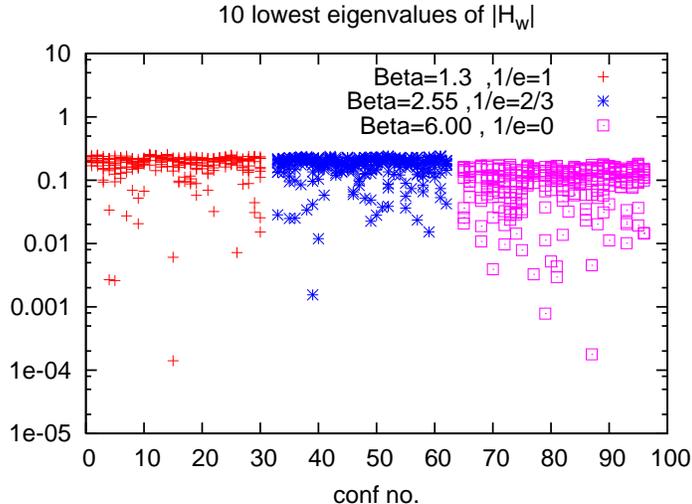}
  \caption{
    Ten lowest eigenvalues of $|aH_W|=|\gamma_5(aD_W-1.6)|$ 
    for gauge configurations with $r_0/a \simeq 5.3$.
    Data are shown for $1/\epsilon$ = 1 (pluses), 2/3
    (stars), and 0 (squares).
    The lattice size is 20$^4$.
  }
  \label{fig:HwL20coarse}
\end{figure}

Figure~\ref{fig:HwL20coarse} shows a typical comparison of
the eigenvalue distribution for three values of
$1/\epsilon$ on a 20$^4$ lattice.
The $\beta$ value is chosen such that the Sommer scale
$r_0/a$ is roughly equal to 5.3, which corresponds to
$a\simeq 0.1$~fm.
From the plot we observe that the density of the low-lying
modes is relatively small for larger values of $1/\epsilon$.
To quantify this statement we list the probability,
$P(<0.1)$, to find the eigenvalue smaller than 0.1 in
Table~\ref{tab:Hw}.
For the above example, the probability is 41\% for
the standard Wilson gauge action ($1/\epsilon$ = 0), but it
decreases to 15\% (9\%) for $1/\epsilon$ = 2/3 (1).
For another lattice spacing ($r_0/a\simeq$~6.5) and lattice
size $16^4$, a similar trend can be found.
In Table~\ref{tab:Hw} we also summarize the ensemble average
of the lowest eigenvalue $\lambda_{\mathrm{min}}$ and
the inverse of condition numbers
$\lambda_{\mathrm{max}}/\lambda_{\mathrm{min}}$ and
$\lambda_{\mathrm{max}}/\lambda_{\mathrm{10}}$, where
$\lambda_{\mathrm{10}}$ and $\lambda_{\mathrm{max}}$ denote
the 10th and the highest eigenvalues respectively.
We may conclude that the lowest eigenvalue is higher in
average for larger $1/\epsilon$.

\begin{table}[tbp]
\begin{center}
\begin{tabular}{cccccccc}
\hline\hline
lattice size & $1/\epsilon$ & $\beta$ & $r_0/a$ & 
$P(<0.1)$ &
$\lambda_{\mathrm{min}}$ &
$\lambda_{\mathrm{min}}/\lambda_{\mathrm{max}}$&  
$\lambda_{\mathrm{10}}/\lambda_{\mathrm{max}}$  \\
\hline
$20^4$ & 1 & 1.3  & 5.240(96)  &  0.090(14) & 0.0882(84) & 0.0148(14) & 0.03970(29)\\
    &  2/3 & 2.55 & 5.290(69)  &  0.145(12) & 0.0604(53) & 0.0101(08) & 0.03651(27)\\
    &    0 & 6.0  & [5.368(22)]&  0.414(29) & 0.0315(57) & 0.0059(34) & 0.02766(46)\\
    &    1 & 1.42 & 6.240(89)  &  0.031(10) & 0.168(13)  & 0.0282(21) & 0.04765(32)\\
    &  2/3 & 2.7  & 6.559(76)  &  0.019(18) & 0.151(11)  & 0.0251(19) & 0.04646(37)\\
    &    0 & 6.13 & [6.642(--)]&  0.084(14) & 0.0861(83) & 0.0126(15) & 0.03775(50)\\
$16^4$ & 1 & 1.3  & 5.240(96)  &  0.053(13) & 0.111(12)  & 0.0187(21) & 0.04455(31)\\
    &  2/3 & 2.55 & 5.290(69)  &  0.067(13) & 0.1038(98) & 0.0174(16) & 0.04239(36)\\
    &    0 & 6.0  & [5.368(22)]&  0.130(20) & 0.0692(90) & 0.0116(15) & 0.03451(62)\\
    &    1 & 1.42 & 6.240(89)  &  0.007(5)  & 0.219(13)  & 0.0367(21) & 0.05233(26)\\
    &  2/3 & 2.7  & 6.559(76)  &  0.020(8)  & 0.191(12)  & 0.0320(19) & 0.05117(29)\\
    &    0 & 6.13 & [6.642(--)]&  0.030(10) & 0.139(10)  & 0.0232(17) &  0.04384(38)\\

\hline
\end{tabular}
\end{center}
\caption{
  The probability $P(<0.1)$ to find the eigenvalue lower
  than 0.1 for the hermitian Wilson-Dirac operator
  $|aH_W|=|\gamma_5(aD_W-1.6)|$.
  Ensemble averages of the lowest eigenvalue and the
  inverse of condition numbers are also listed.
  The Sommer scale $r_0/a$ is the results for $L=16$ lattices.
  The values with [] are from \cite{Necco:2001xg} with an
  interpolation in $\beta$.
}
\label{tab:Hw}
\end{table}

\subsection{Numerical cost}
In the numerical implementation of the overlap-Dirac
operator one often subtracts the low-lying eigenmodes of
$aH_W$ and treats them exactly.
The rest of the modes are approximated by some polynomial or
rational functions.
The numerical cost to operate the overlap-Dirac operator is
dominated by the polynomial/rational part, because the
subtraction have to be done only once for a given
configuration.
Here, we assume that 10 lowest eigenmodes are subtracted and
compare the relative numerical cost on the gauge
configurations with different values of $1/\epsilon$.

\begin{figure}[tbp]
  \centering
  \includegraphics[width=8cm]{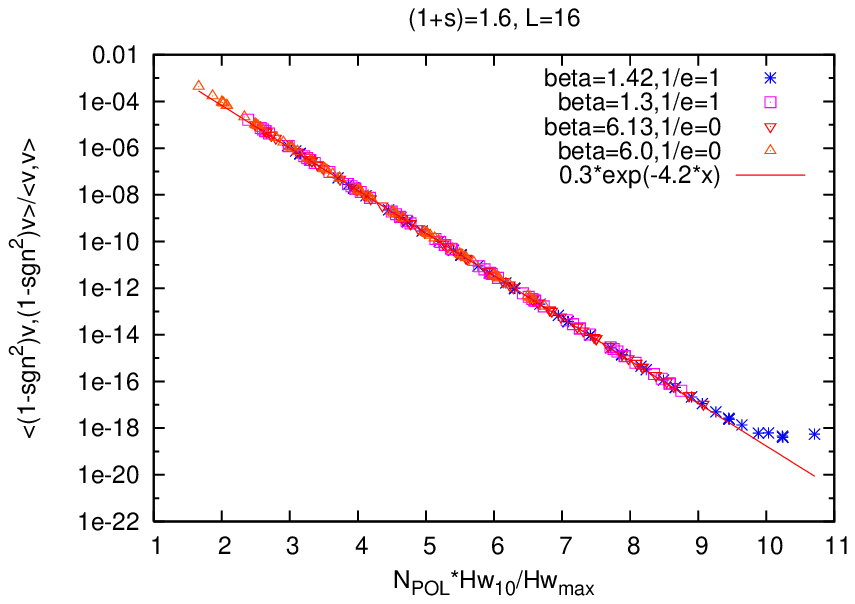}
  \includegraphics[width=8cm]{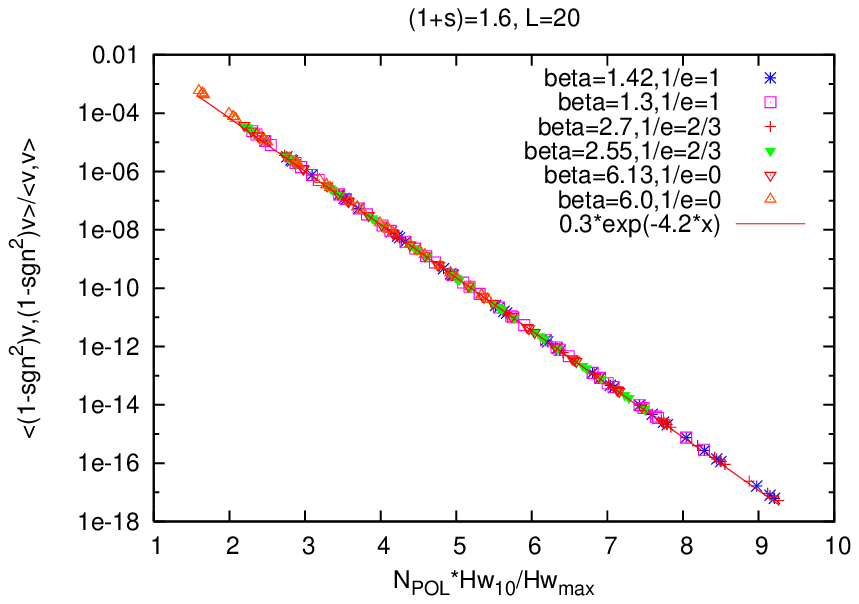}
  \caption{
    The accuracy (\ref{eq:Chebacc2}) as a function of
    $N_{\mathrm{pol}}/\kappa$ for $L$ = 16 (left) and $L$ =
    20 (right).
    We use 4 gauge configurations and 10 values of
    $N_{\mathrm{pol}}$ = 60--195 for each parameter set.
  }
  \label{fig:ABdeterm}
\end{figure}

The accuracy of the Chebyshev polynomial approximation 
$\mbox{sgn}_{\mathrm{Cheb}}(aH_W)$ 
with a degree $N_{\mathrm{poly}}$ can be expressed as
\cite{Giusti:2002sm}
\begin{equation}
  \label{eq:Chebacc2}
  \frac{\langle v |(1-\mbox{sgn}_{\mathrm{Cheb}}^2(aH_W))^2|v \rangle}
  {\langle v|v\rangle}  
  \sim A\exp (-BN_{\mathrm{poly}}/\kappa),
\end{equation}
for a random noise vector $|v\rangle$.
$A$ and $B$ are constants.
We find that they are $A\sim$ 0.3 and $B\sim$ 4.2 almost
independent of the lattice parameters as shown in 
Figure~\ref{fig:ABdeterm}.
The reduced condition number
$\kappa=\lambda_{\mathrm{max}}/\lambda_{\mathrm{10}}$ 
enters in the formula with a combination
$N_{\mathrm{poly}}/\kappa$.
Therefore, the numerical cost, which is proportional to
$N_{\mathrm{poly}}$, depends linearly on $\kappa$ if one
wants to keep the accuracy for the sign function.
From Table~\ref{tab:Hw} we observe that the reduced
condition number is about a factor 1.2--1.4 smaller for
$1/\epsilon$ = 1 than that for the standard Wilson gauge
action.

\begin{figure}[tbp]
  \centering
  \includegraphics[width=8cm]{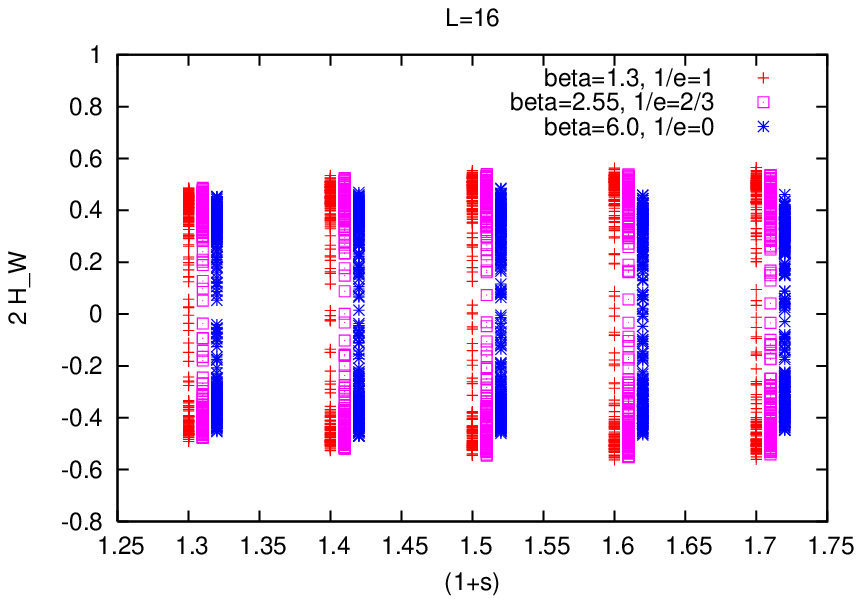}
  \includegraphics[width=8cm]{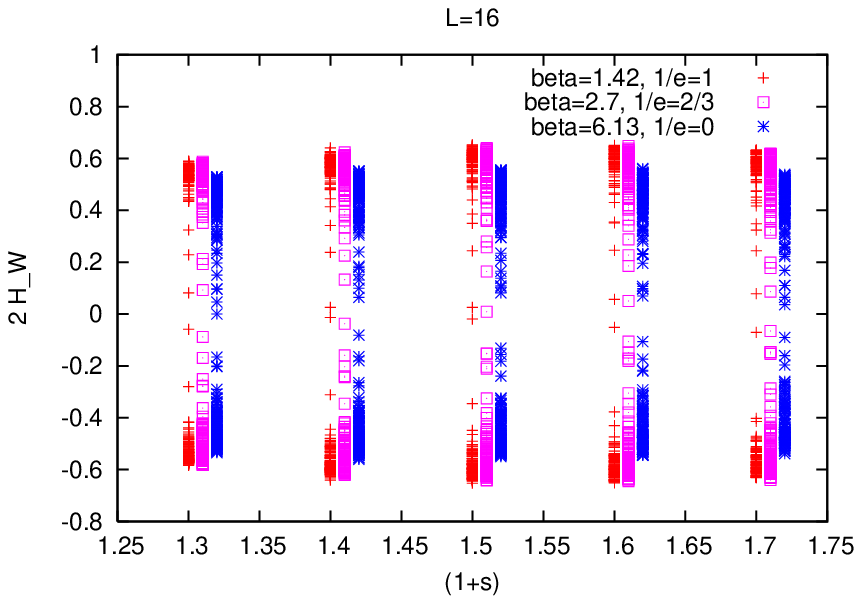}
  \caption{
    Ten near-zero eigenvalues for $r_0/a\sim 5.3$
    (left) and $r_0/a\sim 6.5$ (right).
    Results are plotted as a function of $1+s$.
    Data for $1/\epsilon$ = 2/3 and 0 are slightly shifted
    for clarity.
}
\label{fig:flow}
\end{figure}

We also check that the above observation does not change by
varying the value of $s$ in a reasonable range.
Figure~\ref{fig:flow} shows a typical distribution of the
low-lying eigenmodes for $s$ = 0.2--0.7.
We find that the advantage of the topology conserving gauge
action does not change.
Also, from these plots we can see that
$s\sim 0.6$ is nearly optimal for all cases.

\subsection{Locality}
\begin{figure}[tbp]
  \centering
  \includegraphics[width=8cm]{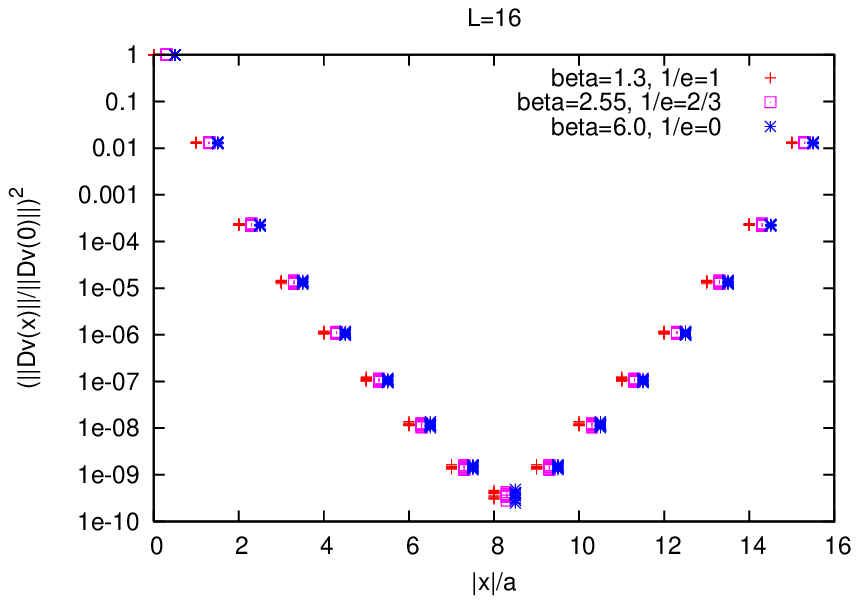}
  \includegraphics[width=8cm]{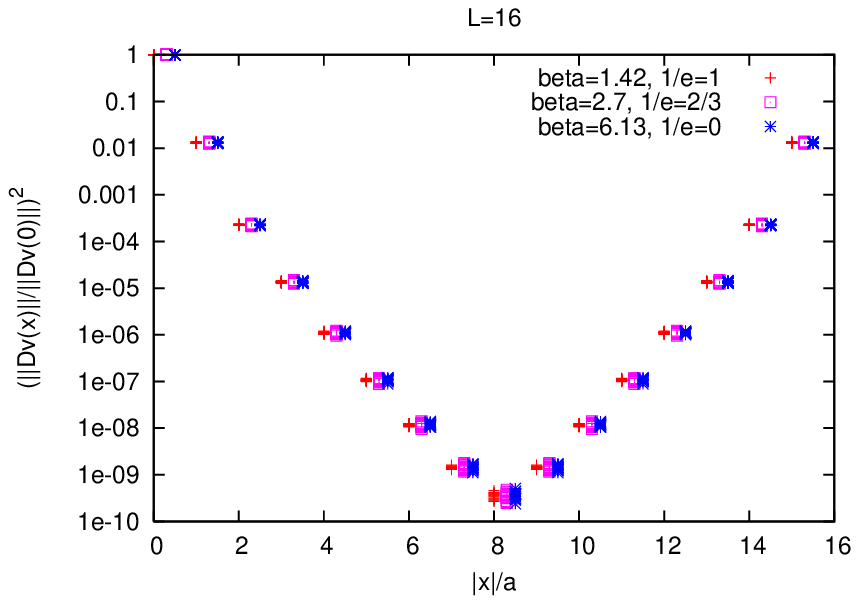}
  \caption{
    $(||D(x,y)v(y)||/||D(0,y)v(y)||)^2$ with $x_0=0$
    measured on 10 gauge 
    configurations for different values of $1/\epsilon$.
    The lattice scale is
    $r_0/a\simeq 5.3$ (left) and 6.5 (right).
  }
  \label{fig:locality}
\end{figure}

If the overlap-Dirac operator is local, the norm 
$||D(x,y)v(y)||$ with a point source vector $v$ at $x_0$
should decay exponentially as a function of $x-x_0$
\cite{Hernandez:1998et}  
\begin{equation}
  ||D(x,y)v(y)|| \sim C \exp (-D|x-x_0|)
\end{equation}
with constants $C$ and $D$.
This behavior is actually observed in
Figure~\ref{fig:locality}.
The plots are shown for different values of $1/\epsilon$ at
the lattice scales $r_0/a\simeq$ 5.3 (left) and 6.5 (right).
We find no remarkable difference on the locality for
different gauge actions. 

Recently, it has been pointed out that the mobility edge is
the crucial quantity which governs the locality of the
overlap-Dirac operator 
\cite{Golterman:2003qe,Golterman:2004cy,Golterman:2005fe,Svetitsky:2005qa}.
It would be interesting to see the dependence of the
mobility edge on the parameters in the topology conserving
action, which is left for future works.

\section{Conclusions}
\label{sec:conclusion}

We study the properties of the topology conserving gauge
action (\ref{eq:admiaction}) in the quenched approximation. 
For small $\epsilon$ ($\lesssim$ 1/20), the parameter to
control the admissibility of the plaquette variable, it is
theoretically known that the topology change is strictly
prohibited, but we investigate the action with 
$\epsilon\sim O(1)$ for the use of practical purposes.
With the (quenched) Hybrid Monte Carlo updation, we find
that the topology change is strongly suppressed for
$1/\epsilon$ = 2/3 and 1, compared to the standard Wilson
gauge action.
The topological charge becomes more stable for fine
lattices, and it is possible to preserve the topological
charge for O(1,000) HMC trajectories at $a\simeq$ 0.08~fm
and $L\simeq$~1.3~fm.
In the same parameter region, the standard Wilson gauge
action changes the topological charge every $O(10)$
trajectories. 
The action is therefore proved to be useful to accumulate
gauge configurations in a fixed topological sector.

We measure the heavy quark potential with this gauge action
at $1/\epsilon$ = 2/3 and 1.
The lattice spacing is determined from the Sommer scale
$r_0$.
With these measurements we also investigate the scaling
violation for short and intermediate distances.
The probe in the short distance is the violation of the 
rotational symmetry, and a ratio $r_c/r_0$ of two different
scale can be used for the intermediate distances.
For both of these we find that the size of the scaling
violation is comparable to the standard Wilson gauge action,
which is consistent with the expectation that the term with
$1/\epsilon$ introduces a difference at most $O(a^4)$.
The action (\ref{eq:admiaction}) shows no disadvantage as
far as the scaling is concerned.

The perturbative expansion of the coupling and Wilson-loops
is available in the literature for general one-plaquette
action.
We write down the coefficients for our particular action
(\ref{eq:admiaction}) and observe that the convergence is
very good if the mean field improvement is applied.
The coupling constant in a certain scheme at a given scale
is consistent among different values of $1/\epsilon$.

As a result of the (approximate) topology conservation, the
low-lying eigenvalues of the Wilson-Dirac operator in the
negative mass regime is suppressed.
This is an advantage in the construction of the
overlap-Dirac operator, since the numerical cost to evaluate
the sign function is proportional to the inverse of the
lowest eigenvalue for a given gauge configuration.
In this case, the gain is about a factor 2--3 at the same
lattice spacing compared to the standard Wilson gauge
action.
If the first several eigenmodes are subtracted and treated
exactly, the gain is marginal, 20--40\%.
Similar improvements have been observed with
the improved gauge actions, such as the
L\"uscher-Weisz, Iwasaki and DBW2.

\section*{ACKNOWLEDGMENTS}\label{sec:acknowlegments}
We thank W.~Bietenholz, L.~Del~Debbio, L.~Giusti,
M.~Hamanaka, T.~Izubuchi, K.~Jansen, H.~Kajiura, 
M.~L\"uscher, H.~Matsufuru, S.~Shcheredin, and T.~Umeda for
discussions. 
HF and TO thank the Theory Group of CERN for the warm
hospitality during their stay.
Numerical works are mainly done on NEC SX-5 at Research
Center for Nuclear Physics, Osaka University.

\end{document}